# Self-Attention Generative Adversarial Network for Iterative Reconstruction of CT Images

**Ruiwen Xing**

University of Washington Bothell

18115 Campus Way NE, Bothell, WA

ruiwen@uw.edu

**Thomas Humphries\***

University of Washington Bothell

18115 Campus Way NE, Bothell, WA

thumphri@uw.edu

**Dong Si\***

University of Washington Bothell

18115 Campus Way NE, Bothell, WA

dongsi@uw.edu

\* Corresponding Authors

## Abstract

*Computed tomography (CT) uses X-ray measurements taken from sensors around the body to generate tomographic images of the human body. Conventional reconstruction algorithms can be used if the X-ray data are adequately sampled and of high quality; however, concerns such as reducing dose to the patient, or geometric limitations on data acquisition, may result in low quality or incomplete data. Images reconstructed from these data using conventional methods are of poor quality, due to noise and other artifacts. The aim of this study is to train a single neural network to reconstruct high-quality CT images from noisy or incomplete CT scan data, including low-dose, sparse-view, and limited-angle scenarios. To accomplish this task, we train a generative adversarial network (GAN) as a signal prior, to be used in conjunction with the iterative simultaneous algebraic reconstruction technique (SART) for CT data. The network includes a self-attention block to model long-range dependencies in the data. We compare our Self-Attention GAN for CT image reconstruction with several state-of-the-art approaches, including denoising cycle GAN, CIRCLE GAN, and a total variation superiorized algorithm. Our approach is shown to have comparable overall performance to CIRCLE GAN, while outperforming the other two approaches.*

# 1. Introduction

Computed tomography (CT) imaging is a medical imaging modality in which an image of the patient anatomy is reconstructed from X-ray data gathered from a number of angles around the patient. Unlike planar X-ray scans, in which anatomical structures are superimposed in a two-dimensional projection, CT provides a fully three-dimensional image of the body. The image can be volume-rendered or, more typically, viewed as two-dimensional slices in the transverse, sagittal or coronal planes. Due to its high resolution and ability to provide 3D information, CT is highly effective for detecting and localizing tumors, internal hemorrhages, embolisms, and other serious medical conditions; it is estimated that roughly 80 million CT scans were performed in the United States in 2017 [1].

While CT image reconstruction is used every day in clinical practice with excellent results, a number of challenging problems still remain. Concern about the dose associated with repeated CT scans has led to considerable interest in *low-dose imaging* [2] in recent years. Low-dose imaging can be achieved by reducing the intensity of the X-ray beam. Because X-ray measurements are noisy, however, (typically modeled as following a Poisson process), the signal to noise ratio (SNR) of the measurement decreases with the measured intensity. Reducing beam intensity therefore produces noisier measurements, and reduces image quality.

A second way of reducing dose is to acquire fewer X-ray measurements, while maintaining the same beam intensity. For example, to reduce the number of measurements by a factor of 10, the angular increment between successive views can be increased by the same factor, by rapidly switching the source on and off. While this preserves the signal-to-noise ratio of individual measurements, the X-ray transform of the image may no longer be adequately sampled. This results in ray-like artifacts radiating from the center of the object. This type of imaging is typically known as *sparse-view imaging* [3].

Finally, to accurately reconstruct an image, one typically requires measurements acquired from at least 180° around the patient for parallel beam geometries, and somewhat more for fan-beam geometries. In *limited-angle imaging* [3], it may only be possible to acquire data along a shorter arc, resulting in a contiguous section of "missing" angular measurements. Due to the lack of information about edges of structures tangent to the missing rays, images reconstructed from limited-angle data suffer from blurred and missing edges along the corresponding directions.

These challenging problems can be addressed in in a number of ways. For low-dose imaging, several hardware-based solutions are possible. These include automatic exposure control [1][4][5] (lowering beam intensity for measurements passing through less material), automatic tube potential selection to enhance contrast [6], and dynamic z-axis collimators to reduce so-called overscanning in helical CT [7][8]. These solutions do not generalize to sparse-view or limited-angle imaging, however, and typically require specialized equipment or camera software.

Low-dose imaging can also be achieved by pre-processing or filtering the sinogram with software-based approaches [9-11]. Alternatively, low-dose, sparse-view, or limited-angle images can be reconstructed using iterative methods which incorporate prior information to regularize the solution. Approaches incorporating smoothing or sparsifying transforms, such as total variation, have been particularly successful [3,12-15].

Over the last few years, there has been a dramatic increase in interest in the application of techniques from deep learning, such as convolutional neural networks (CNNs), to address these problems in CT imaging. This interest is motivated in large part by the recent success of deep learning in addressing problems in other areas of image processing, such as classification and segmentation. A special issue of *IEEE Transactions on Medical Imaging* published in 2018 highlights a number of promising approaches [16]. Broadly speaking, the approaches can be separated into three categories: using the CNN as a post-processing step to denoise or remove artifacts from images reconstructed using conventional methods; incorporating the CNN into an iterative algorithm as a means of incorporating data-driven prior information, or fully learning the inversion to reconstruct the image from the sinogram.

Post-processing approaches have been primarily applied to low-dose image denoising[17].

Such an approach can be enhanced by applying CNNs in the wavelet domain [18] or by using more sophisticated network structures, such as autoencoders [19] or generative adversarial networks (GANs) [20][21]. Sparse-view and limited angle imaging have similarly been treated by using a CNN as a post-processing step[22], using a variety of CNN architectures such as DenseNet[23] and wavelet-domain U-Nets [24][25]. Iterative algorithms incorporating CNNs include the LEARN method [26], applied to sparse-view imaging; PWLS-ULTRA [27] for low-dose imaging; and the CNN-based projected gradient approach of [28]. Examples of fully-learned inversion methods include the limited-angle reconstruction method of [29] and the iCT-Net method[30], which was applied to low-dose, sparse-view, and interior tomography problems.

With the exception of iCT-Net[30], most of the above approaches were trained on a particular low dose, sparse view or limited angle scenario. Given the computational effort required to train deep neural networks, it is of considerable interest to investigate whether a single network can be trained to improve image quality under a variety of imaging scenarios. Inspired by the OneNet architecture [31], we have developed a convolutional autoencoder-based generative adversarial network for CT imaging denoising and reconstruction in this paper. The network was trained on a mixed dataset of low-dose, sparse-view, and limited-angle data, and employed as a signal prior between iterations of a well-known simultaneous algebraic reconstruction technique (SART) for CT. Additionally, we incorporated the concept of self-attention[32,33] to utilize long-range dependency relationships in the reconstructed image. In numerical experiments, we have compared the approach with two other CNN-based reconstruction algorithms, as well as a total variation "superiorized"[14] reconstruction approach. All of our code is publicly available on GitHub at https://github.com/TDHumphries/SART-SAGAN/.

## 2. Methods

### 2.1. Iterative algorithm

Iterative algorithms for CT image reconstruction are typically based on the linear model $\mathbf{Ax} = \mathbf{b}$, where $\mathbf{x}$ is the image to be reconstructed, $\mathbf{b}$ is the vector of measured data, and $\mathbf{A}$ is the CT system matrix. One well-known method is the simultaneous algebraic reconstruction technique (SART) [33] – [37]. First, we define diagonal matrices $\mathbf{D}$ and $\mathbf{M}$ by:

$$\mathbf{D}_{k,k} = \frac{1}{\sum_{i=1}^{J}|a_{i,k}|} \, ; \, \mathbf{M}_{j,j} = \frac{1}{\sum_{i=1}^{K}|a_{j,i}|} \, ; \tag{1}$$

that is, $\mathbf{D}$ is a $k \times k$ matrix whose diagonal entries are the reciprocals of the column sums of $\mathbf{A}$, and $\mathbf{M}$ is a $j \times j$ matrix whose entries are the reciprocals of the row sums. Second, we partition the sinogram into $N_w$ equally spaced subsets, indexed by $w$; for example, if $N_w$=12, then $w$=1 consists of the first, thirteenth, twenty-fifth, etc. columns of the sinogram, $w$=2 to the second, fourteenth, twenty-sixth, etc. This so-called block iterative approach accelerates the convergence of the algorithm. A full iteration of this block-iterative approach can then be described as:

$$\mathbf{x}^{(i+1)} = P_{\Omega}\left(B_{N_w} \dots B_2 B_1(\mathbf{x}^{(i)})\right) \tag{2}$$

where,

$$B_w(\mathbf{x}) = \mathbf{x} - \mathbf{D}_w(\mathbf{A}_w)^T \mathbf{M}_w(\mathbf{A}_w \mathbf{x} - \mathbf{b}_w) \tag{3}$$

The subscript $w$ indicates that only the rows of $\mathbf{A}$ and $\mathbf{b}$ corresponding to subset $w$ are used, including when forming the matrices $\mathbf{D}_w$ and $\mathbf{M}_w$. The operation $P_{\Omega}$ represents projection onto the feasible set of solutions ($\Omega$); for example, images whose components are all non-negative. This iteration can be shown to converge to a weighted least squares solution of the problem, provided that such a solution lies within $\Omega$ [35-36]. In cases of noisy or under sampled data, however, this solution may not be desirable, which necessitates the use of prior information in the reconstruction.

### 2.2. Generative Adversarial Network for reconstruction

In this paper, we propose combining iterative CT reconstruction with prior information based on a single pre-trained CNN to reconstruct sparse-view, low-dose, and limited-angle images. Our approach is inspired by the One-Net algorithm[31], which proposed training a single network as a "quasi-projector" onto a given space of images, then pairing it with an iterative algorithm to solve arbitrary linear inverse problems. The authors trained the network on photographic images, then combined it with the alternating direction method of multipliers (ADMM) method to solve inpainting, super-resolution, and compressive-sensing problems. In our work we train the network on CT images, and use SART as the reconstruction algorithm, as it is more widely used for CT image reconstruction.

The conceptual framework is shown in Fig. 1. The system consists of a generative adversarial network (GAN) paired with a SART iteration. The GAN module consists of two components: a generator, which attempts to recover a clean image from one containing noise and artifacts, and a discriminator, which attempts to distinguish between the images produced by the generator and real high-quality CT images. As a generative model, GAN is designed to create new data based on information provided, rather than remove or extract data [38]. This makes it suitable for image reconstruction tasks.

After training, the generator network is integrated into the SART algorithm. The generator is applied between iterations of SART as a quasi-projector (essentially replacing the operator $P_\Omega$ in (2)), to guide the reconstruction process towards a higher-quality image.

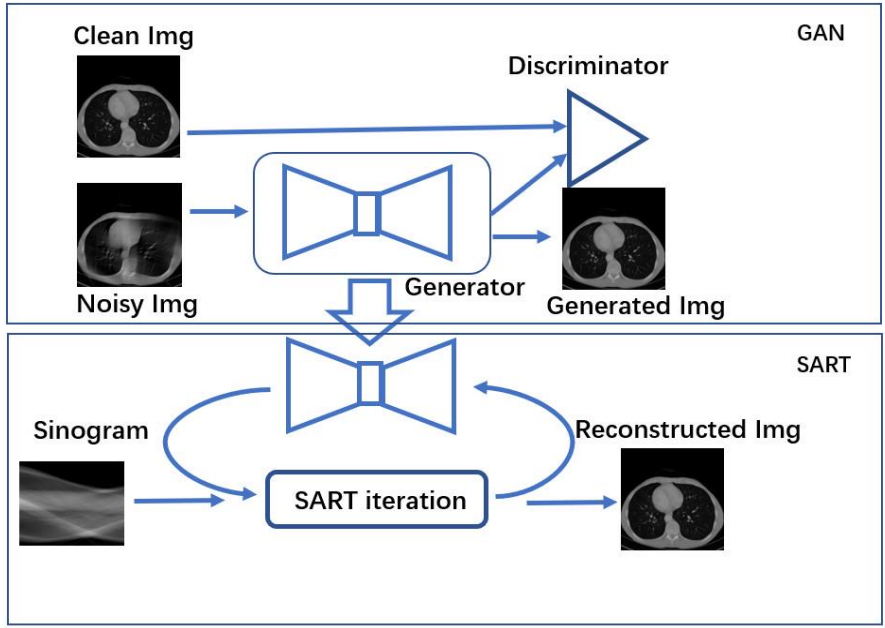

**Fig. 1** General structure of proposed method. The GAN component (top half) is trained on paired data consisting of clean and "noisy" (low-dose, sparse-view, or limited angle) data. The trained generator is then incorporated into the SART algorithm (bottom half)

The structure of the generator network is based on a convolutional autoencoder, consisting of ten convolutional (encoding) layers and ten deconvolutional (decoding) layers. During the encoding process, the size of the image is decreased and the color dimension of image is increased. As shown in Fig. 2, after the encoder, a self-attention block [32,33] is added to increase the ability of generator to recognize long range dependencies in the image. To control the amount of data in the attention map, we assign an attention to each nine adjacent pixels. Average pooling is used to generate self-attention.

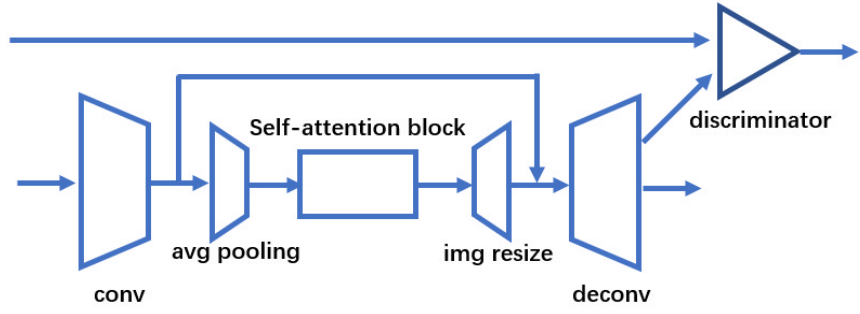

**Fig. 2** Structure of the generative adversarial network shown in Fig. 1

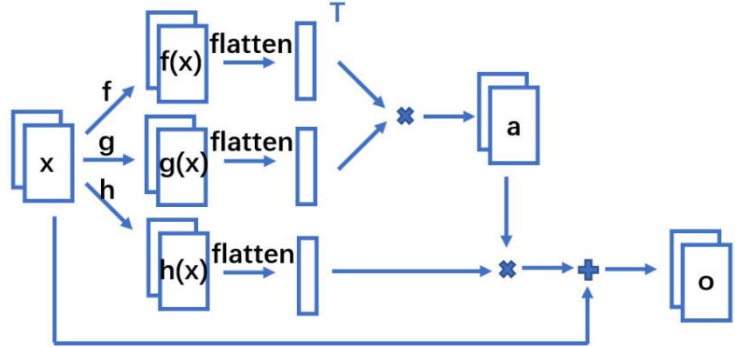

**Fig. 3** Structure of the self-attention block shown in Fig. 2

Fig. 3 (adapted from [32][33]) shows the structure of the self-attention block. The self-attention block transforms the image feature tensor $x$ from the previous coded layer, having dimensions $(m \times n \times c)$, into two feature spaces $f$ and $g$. Here, $m$ and $n$ represent the number of rows and columns in $\mathbf{x}$, and $c$ is the number of channels. The feature spaces are generated as

$$f(\mathbf{x}) = \mathbf{W}_f\mathbf{x}, \tag{4}$$
$$g(\mathbf{x}) = \mathbf{W}_g\mathbf{x}, \tag{5}$$

where $\mathbf{W}_f$ and $\mathbf{W}_g$ are trainable weighting matrices.

The attention map, $\mathbf{a}$, is an $(mn) \times (mn)$ matrix, whose $(i,j)$th entry represents the attention on the $j^{th}$ image feature with respect to the the $i^{th}$ image feature. The map is calculated as

$$\mathbf{a} = softmax\left(flatten(f(\mathbf{x}))^T flatten(g(\mathbf{x}))\right), \tag{6}$$

where the *flatten* function reshapes $f(\mathbf{x})$ and $g(\mathbf{x})$ to two $(mn) \times c$ matrices. The output, $o$, of the self-attention block is then calculated as

$$\mathbf{o} = \gamma \times reshape(flatten(h(\mathbf{x}))\mathbf{a}) + \mathbf{x} \tag{7}$$

where $\gamma$ is a learnable scalar, and

$$h(\mathbf{x}) = \mathbf{W}_h\mathbf{x}, \tag{8}$$

where $\mathbf{W}_h$ is a trainable weighting matrix with the same dimensions as $\mathbf{W}_f$ and $\mathbf{W}_g$. The *reshape* function converts the product of the flattened $h(x)$ with $a$ back to a tensor of size $m \times n \times c$. This process generates a set of attention weights for each pixel block, allowing the network to model long range dependencies in the image. As indicated in Fig. 2, we also include a shortcut across the self-attention block, to preserve information from the previous layer.

### 2.3. Training

The training data set was based on 3900 CT image slices downloaded from the Cancer Imaging Archive[1]. The images were taken from a lung study and included anatomical regions ranging from the lower abdomen to upper thorax. The dataset was augmented fourfold by rotating the images. The ASTRA toolbox [39][40] was then used to generate sinogram data from the images, corresponding to a number of different imaging scenarios:

1. Normal-dose sinograms: 180 degrees of data, 900 views, initial beam intensity of $I_0 = 10^6$ counts per measurement.
2. Low-dose images: beam intensity lowered to $I_0 = 10^5$ and $10^4$.
3. Sparse-view sinograms: number of views lowered to 100 views and 50 views.
4. Limited-angle sinograms: angular range reduced to 140 degrees (700 views).

These values were chosen based on those used in the literature for simulating low-dose, sparse-view and limited-angle CT, e.g. [19],[25-27]. In all cases, Poisson noise was added to the sinograms proportionally to the beam intensity. From each sinogram, we then reconstructed images using 20 iterations of the classical SART method, with 50 subsets. Finally, the normal-dose reconstructed images were paired with corresponding low-dose, sparse-view or limited angle images and fed into the GAN for training.

The GAN was trained using the Adaptive Moment Estimation (Adam) [41] optimizer. During training the learning rate was gradually reduced from 0.0001 to 0.000002, to promote convergence of the optimizer. The generator and discriminator losses were given, respectively, by

$$loss_G = \parallel \mathbf{Y} - G(\mathbf{X}) \parallel_2^2 + \left( \sum_{i=0}^{h} \sum_{j=0}^{w} \left( G(\mathbf{X}_{i,j}) - G(\mathbf{X}_{i-1,j}) \right) \right)^2 + \sum_{i=0}^{h} \sum_{j=0}^{w} \left( G(\mathbf{X}_{i,j}) - G(\mathbf{X}_{i,j-1}) \right) \right)^2 \quad (9)$$

$$loss_D = \parallel D(\mathbf{Y}) - 1 \parallel_2^2 + \parallel D(G(\mathbf{X})) - 0 \parallel_2^2 \quad , \quad (10)$$

where $\mathbf{Y}$ represents the normal-dose image and $\mathbf{X}$ the low-dose, sparse-view or limited angle image, $G(\mathbf{X})$ the output from the generator, and $D(\mathbf{Y})$ the output from the discriminator. The discriminator output is 1 if the image is believed to be a genuine low-dose image, and 0 if it believed to be output form the generator. The generator loss consists of $l_2$ error between the normal-dose image and the generator output, as well as a smooth loss term in order to reduce noise in the generated images.

The progress of the generator loss and discriminator loss during training is shown in Fig. 4. Training was stopped when the two loss functions reached a steady state, after roughly $2.5 \times 10^5$ training iterations. The average peak signal-to-noise ratio (PSNR) of the validation set during training is shown in Fig. 5. We use PSNR (described in the next section) as one metric of image quality in our numerical experiments. Fig. 5 indicates that the choice of loss functions (9)(10) is effective in training the network to improve image quality.

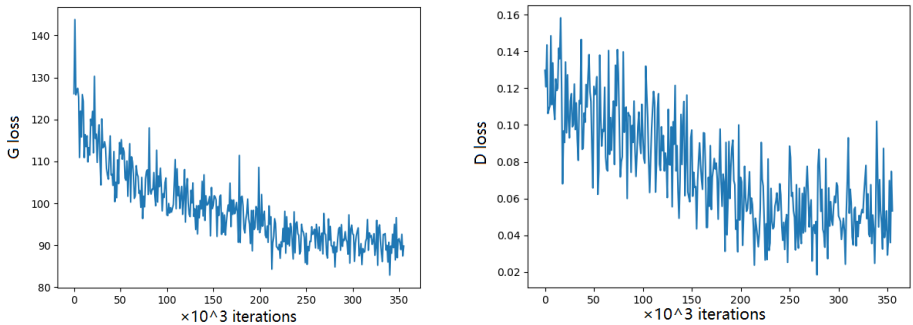

**Fig. 4** Generator loss (left) and discriminator loss (right) during training of the GAN

---

[1] https://www.cancerimagingarchive.net/

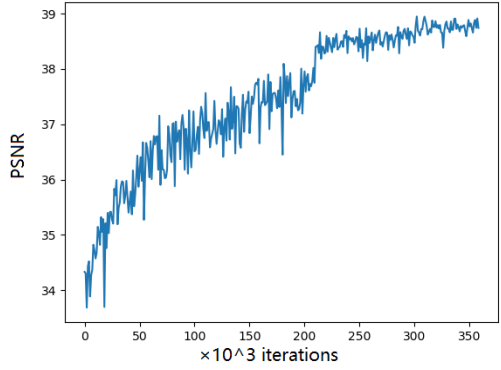

**Fig. 5** PSNR of validation dataset achieved during training

## 2.4. Evaluation methods

We evaluated the quality the images produced by our method by comparing them with the corresponding images reconstructed by SART using normal dose data. The two metrics we used are Peak Signal-to-Noise Ratio (PSNR) and Structural SIMilarity (SSIM).

Peak Signal-to-Noise Ratio (PSNR) measures the ratio between the maximum possible value (or power) of an image, versus the power of the image noise. It is defined as

$$PSNR = 10 \cdot log_{10}\left(\frac{MAX_y^2}{MSE}\right) \tag{11}$$

where $MAX_y$ is the maximum intensity value of the ground truth image $y$, and the mean squared error (MSE) is given by:

$$MSE = \frac{1}{mn}\sum_{i=0}^{m-1}\sum_{j=0}^{n-1}[\mathbf{x}(i,j) - \mathbf{y}(i,j)]^2 \tag{12}$$

The PSNR is measured in decibels; the higher the PSNR, the better the (pixelwise) agreement between $x$ and $y$. The value is infinite if the two images are identical.

The structural similarity (SSIM) index [42] is defined as:

$$SSIM(x,y) = \frac{(2\mu_x\mu_y+c_1)(2\sigma_{xy}+c_2)}{(\mu_x^2+\mu_y^2+c_1)(\sigma_x^2+\sigma_y^2+c_2)} \tag{13}$$

where $\mu_x$ is the average pixel value of image $x$, $\mu_y$ is the average pixel value of image $y$, $\sigma_x$ is the variance of image $x$, $\sigma_y$ is the variance of image $y$, $\sigma_{xy}$ is the covariance of image $x$ and image $y$, $c_1$ and $c_2$ are variables to stabilize the division with weak denominator. SSIM is intended to address a well-known shortcoming of PSNR; namely, that images with vastly different perceived quality may have identical MSE. The SSIM value ranges from 0 to 1, with 1 indicating a perfect match between $x$ and $y$.

## 2.5. Experiments

As mentioned in Section 2.3, we trained a single neural network on a dataset consisting of low-dose, sparse view, and limited angle sinogram data. We then assessed the performance of our reconstruction method integrating the network with SART (Fig. 1) on the three types of problem, both visually and using PSNR and SSIM for quantitative evaluation. To test the robustness of the approach, we varied the number of views, angular extent, and noise levels in our test problems to include scenarios on which the network was not trained. For example, while the network was only trained with 140 degree limited angle data, we also tested it on 160 and 120 degree scenarios. The low dose, sparse view and limited angle data were generated from 160 CT images, which were not included in the training set. As with the training data, these were generated by rotating 40 images from the Cancer Imaging Archive, and sinogram data were generated using the ASTRA Toolbox.

We compared our results with two state-of-the-art neural network approaches: denoising cycle GAN [43] and CIRCLE GAN [44]. Details of these two approaches are described in the respective papers. We also compared with a "superiorized" reconstruction approach [14], denoted SART-TV, which incorporates total variation (TV) minimization between each iteration of SART. TV minimization is a standard mathematical approach to low-dose and sparse-view CT imaging. We also compared against the standard ("pure") SART algorithm to provide insight into the improvement offered by the other approaches. To avoid overemphasizing small differences in performance when comparing the approaches, we only consider cases where there is more than a 2% difference between PSNR or SSIM values to represent a meaningful difference in performance.

## 3. Results

### 3.1. Results on limited angle data

Table 1 shows the results of each method on 160, 140, and 120 degrees limited angle data. The result is based on the average PSNR and SSIM over all images in the test set. The proposed method gave better results than all other methods in two of the three cases. Generally, the improvement in performance of our method versus the others increased with the difficulty of the problem (i.e. as angular extent decreased). For example, the improvement in PSNR with respect to SART-TV was 7% with 160 degree data, 18% with 140 degree data, and 15% with 120 degree data. When compared with CIRCLE GAN, the second-best performing method, our approach gave comparable results on 160 degree data, an improvement of 4% on 140 degree data, and 7% on 120 degree data. Averaged over all experiments, the proposed method gave a 4% improvement in PSNR compared to CIRCLE GAN.

The results also indicate that the approaches based on neural networks have better performance on limited angle data than SART-TV. While SART-TV provides some improvement over pure SART for the limited angle scenario, the TV function lacks the ability to effectively correct limited angle artifacts, which are nonlocal phenomena.

Fig. 6 shows a representative image reconstructed from 120 degree data (the most challenging scenario) using each approach. The white arrows indicate the region most badly affected by the missing data. As we can see, the proposed method was most effective in recovering this region.

Table 1. Peak signal-to-noise ratio (PSNR) and structural similarity (SSIM) values for limited angle experiments, averaged over all images in the validation set. Columns correspond to the three limited-angle scenarios, as well as the overall average. Rows correspond to the different methods, with the first row being the PSNR value, and the second row the SSIM value. Best overall PSNR and SSIM values in each column are highlighted in bold, together with any values within 2% of the best value. The percentage difference is indicated in parentheses.

| PSNR (avg) SSIM (avg) | 160 degrees | 140 degrees | 120 degrees | Average |
|---|---|---|---|---|
| Proposed method | **38.6** | **36.7** | **31.8** | **35.7** |
| | **0.986** | **0.981** | **0.968** | **0.978** |
| Denoising cycle GAN | **38.0 (-1.6%)** | 34.7 (-5.4%) | 29.1 (-8.5%) | 33.9 (-5.0%) |
| | **0.985 (-0.1%)** | **0.978 (-0.3%)** | 0.952 (-1.6%) | **0.972 (-0.6%)** |
| CIRCLE GAN | **38.1 (-1.3%)** | 35.1 (-4.3%) | 29.5 (-7.2%) | 34.2 (-4.2%) |
| | **0.980 (-0.6%)** | **0.980 (-0.1%)** | **0.953 (-0.1%)** | **0.971 (-0.7%)** |
| SART-TV | 36.0 (-6.7%) | 31.0 (-15.5%) | 27.7 (-12.9%) | 31.6 (-11.5%) |
| | **0.978 (-0.8%)** | **0.978 (-0.3%)** | 0.942 (-2.7%) | **0.966 (-1.2%)** |
| Pure SART | 34.7 (-10.1%) | 29.9 (-18.5%) | 27.4 (-13.8%) | 30.7 (-14.0%) |
| | 0.964 (-2.2%) | **0.930 (-5.2%)** | 0.901 (-6.9%) | 0.932 (-4.7%) |

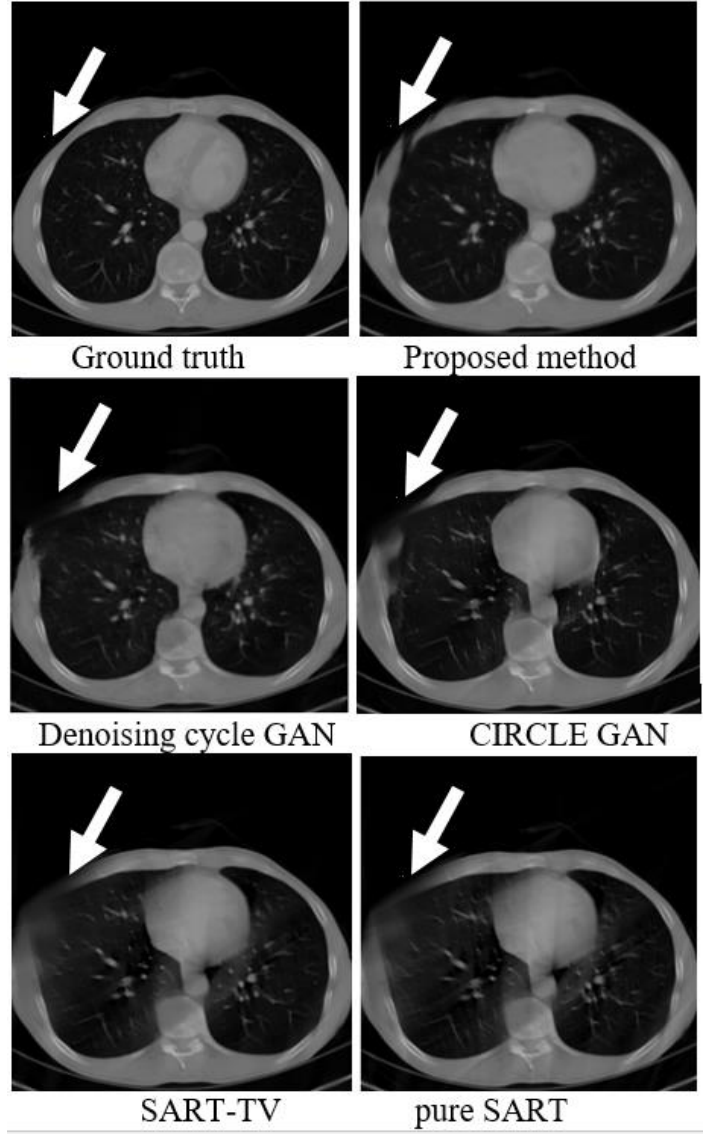

Ground truth      Proposed method

Denoising cycle GAN      CIRCLE GAN

SART-TV      pure SART

**Fig. 6** Image reconstructed from 120 degree limited angle data using the five different approaches, plus the ground truth image for reference

### 3.2. Results on low-dose data

As shown in Table 2, the denoising cycle GAN and CIRCLE GAN methods gave better results than the proposed method in the two lower noise scenarios (X-ray beam intensity $I_0 = 10^6$ and $10^5$ as described in Section 1). At the highest noise level ($I_0 = 10^4$), however, the performance of the proposed method was comparable to CIRCLE GAN, while cycle GAN performed significantly worse. When averaging over all three scenarios, CIRCLE GAN gave an average PSNR that is 2% better than the proposed approach. Fig. 7 shows reconstructed images, zoomed in to show fine detail such as the veins in the lungs.

Table 2. Peak signal-to-noise ratio (PSNR) and structural similarity (SSIM) values for low-dose experiments, averaged over all images in the validation set. Columns correspond to the three limited-angle scenarios, as well as the overall average. Rows correspond to the different methods, with the first row being the PSNR value, and the second row the SSIM value. Best overall PSNR and SSIM values in each column are highlighted in bold, together with any values within 2% of the best value. The percentage difference is indicated in parentheses.

| PSNR (avg) SSIM (avg) | Noise level $I_0 = 10^6$ | Noise level $I_0 = 10^5$ | Noise level $I_0 = 10^4$ | Average |
|---|---|---|---|---|
| Proposed method | 40.0 (-2.9%) | 39.6 (-5.2%) | **35.3** | 38.3 (-2.0%) |
| | **0.989 (-0.2%)** | **0.987 (-0.4%)** | **0.957** | **0.977** |
| Denoising cycle GAN | **41.2** | 40.7 (-2.6%) | 26.6 (-24.6%) | 36.2 (-7.4%) |
| | **0.991** | **0.988 (-0.3%)** | 0.741 (-22.6%) | 0.907 (-7.2%) |
| CIRCLE GAN | **40.8 (-1.0%)** | **41.8** | **34.7 (-1.7%)** | **39.1** |
| | **0.991** | **0.991** | **0.948 (-0.9%)** | **0.977** |
| SART-TV | **40.5 (-1.7%)** | 38.6 (-7.6%) | 33.4 (-5.4%) | 37.5 (-4.1%) |
| | **0.991** | **0.986 (-0.5%)** | **0.945 (-1.2%)** | **0.974 (-0.3%)** |
| Pure SART | 37.1 (-10.0%) | 29.6 (-29.2%) | 20.7 (-41.4%) | 29.1 (-25.6%) |
| | **0.978 (-1.3%)** | 0.968 (-2.3%) | 0.878 (-8.2%) | 0.941 (-3.7%) |

### 3.3. Results on sparse view data

We tested the performance of each approach on sparse-view data using 100, 60, and 50 views. As shown in Table 3, in the 100 view case, the cycle GAN and CIRCLE GAN gave slightly better performance. The proposed method gave comparable results for the 60 view case. Unlike for limited angle imaging, the SART-TV approach was competitive with the neural networks for sparse-view imaging, even providing the best results for the 50-view case. Some reconstructed images from the 60 view case are shown in Fig. 8, zoomed in to show more detail. The characteristic streaking artifacts caused by sparse-view data are still visible in the images reconstructed using cycle GAN and CIRCLE GAN, but have been mostly eliminated by the proposed approach.

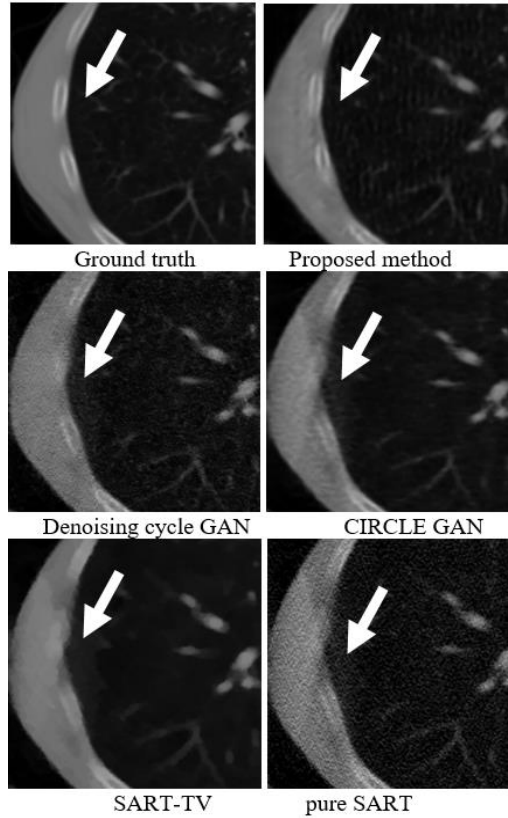

Ground truth     Proposed method

Denoising cycle GAN     CIRCLE GAN

SART-TV     pure SART

**Fig. 7** Image reconstructed from low-dose data (with initial beam intensity $I_0 = 10^4$), using the five different approaches, plus the ground truth image for reference

Table 3. Peak signal-to-noise ratio (PSNR) and structural similarity (SSIM) values for sparse-view experiments, averaged over all images in the validation set. Columns correspond to the three sparse-view scenarios, as well as the overall average. Rows correspond to the different methods, with the first row being the PSNR value, and the second row the SSIM value. Best overall PSNR and SSIM values in each column are highlighted in bold, together with any values within 2% of the best value. The percentage difference is indicated in parentheses.

| PSNR (avg) SSIM (avg) | 100 views | 60 views | 50 views | Average |
|---|---|---|---|---|
| Proposed method | 39.3 (-2.5%) | **38.1** | 36.6 (-3.2%) | **38.0 (-1.3%)** |
| | **0.987 (-0.2%)** | **0.983 (-0.7%)** | **0.974 (-0.2%)** | **0.982 (-0.2%)** |
| Denoising cycle GAN | **40.1 (-0.5%)** | 37.0 (-2.9%) | **37.1 (-1.9%)** | **38.1 (-1.0%)** |
| | **0.988 (-0.1%)** | **0.990** | 0.975 (-0.4%) | **0.984** |
| CIRCLE GAN | **40.3** | **37.8 (-0.8%)** | **37.3(-1.3%)** | **38.5** |
| | **0.989** | **0.981 (-0.9%)** | **0.976 (-0.3%)** | **0.982 (-0.2%)** |
| SART-TV | **39.7 (-1.5%)** | 37.9 (-0.5%) | **37.8** | **38.5** |
| | **0.988 (-0.1%)** | **0.983 (-0.7%)** | **0.979** | **0.984** |
| Pure SART | 38.2 (-5.2%) | 35.4 (-7.1%) | 34.8 (-7.9%) | 36.1 (-6.2%) |
| | **0.972 (-1.7%)** | 0.955 (-3.5%) | 0.946 (-3.3%) | 0.958 (-2.6%) |

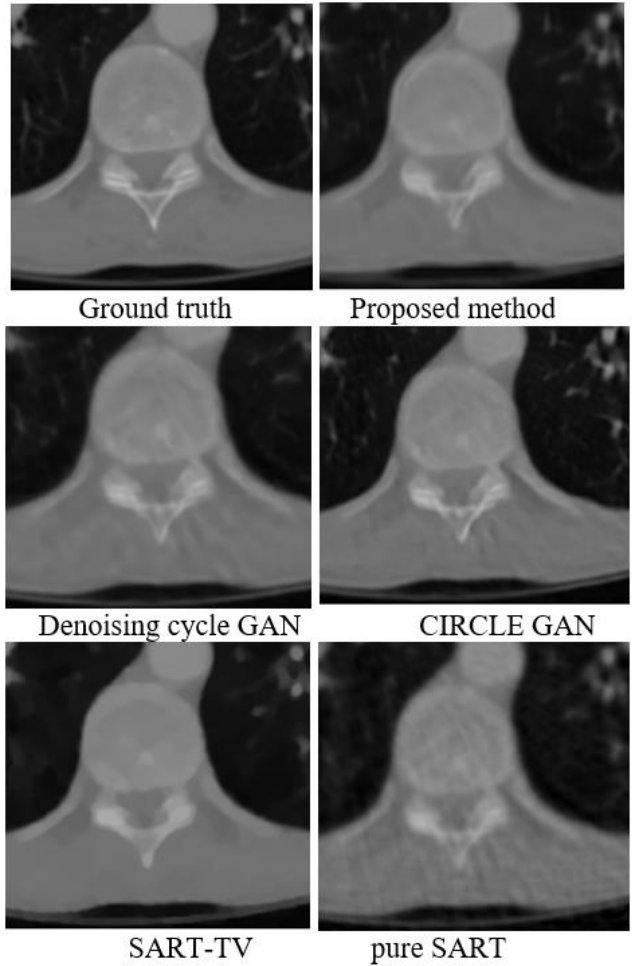

**Fig. 8** Image reconstructed from 60 view sparse view data using the five different approaches, plus the ground truth image for reference

### 3.4. Effect of self-attention block

To test the effectiveness of the self-attention block (Fig. 2 and Fig. 3), we also conducted an ablation study in which the self-attention block was removed from our generator network. That is, we trained the generator network only with encoding and decoding layers, and performed the same tests. The results of the ablation study are shown in Table 4.

Table 4. Peak signal-to-noise ratio (PSNR) and structural similarity (SSIM) evaluation of proposed method with and without self-attention block. Rows correspond to the different imaging scenarios, with the first row being the PSNR value, and the second row the SSIM value. Best overall PSNR and SSIM values in each row are highlighted in bold, together with any values within 2% of the best value. The percentage difference is indicated in parentheses.

| Scenario | Group | With SA | Without SA |
|---|---|---|---|
| Limited angle | 160 degrees | **38.6** | 37.7 (-2.3%) |
| | | **0.986** | **0.981 (-0.5%)** |
| | 140 degrees | **36.7** | **36.4 (-0.8%)** |
| | | **0.980 (-0.1%)** | **0.981** |
| | 120 degrees | **31.8 (-0.1%)** | **32.1** |
| | | **0.968** | **0.968** |
| Low-dose | $I_0 = 10^6$ | **39.8** | 39.0 (-2.0%) |
| | | **0.989** | **0.987 (-0.2%)** |
| | $I_0 = 10^5$ | **39.6** | 38.9 (-1.8%) |
| | | **0.987** | **0.986 (-0.1%)** |
| | $I_0 = 10^4$ | **35.3** | 33.8 (-4.2%) |
| | | **0.957** | 0.912 (-4.7%) |
| Sparse view | 100 views | **39.3** | 38.3 (-2.5%) |
| | | **0.987** | **0.985 (-0.2%)** |
| | 60 views | **38.1** | 36.7 (-3.7%) |
| | | **0.983** | **0.977 (-0.6%)** |
| | 50 views | **36.6** | 35.8 (-2.2%) |
| | | **0.974** | **0.972 (-0.2%)** |

As we can see in Table 4, the model with self-attention block performs better than the model without self-attention in most imaging scenarios, with comparable performance in the others. We conclude that using self-attention enhanced the performance of our neural net, most notably in the sparse-view case.

## 4. Discussion

The results of all the experiments are summarized in Table 5 below. Across all experiments, the average performance of our method is comparable to CIRCLE GAN [44], providing somewhat better performance on limited angle imaging, somewhat worse performance on low-dose imaging, and comparable performance on sparse-view imaging. We achieve an overall improvement of 3% in PSNR values compared to the denoising cycle GAN, 4% compared to SART-TV, and 17% compared to the pure SART algorithm.

Contrasting our approach with CIRCLE GAN, a major difference is that CIRCLE GAN is applied as a post processing step to images reconstructed using conventional methods, while our approach incorporates the neural network into an iterative reconstruction algorithm. There are advantages and disadvantages to each approach. Post-processing is more computationally efficient, because the images themselves can be reconstructed using fast algorithms like filtered back projection, rather than slower iterative approaches. On the other hand, after the reconstructed image is passed through the post-processing step, there is no guarantee that the image produced agrees with the original measured (sinogram) data. Instabilities in the network may therefore result in image artifacts [45]. Integrating the neural network within an iterative algorithm provides a feedback loop (Fig. 1) whereby the output of the network is checked against the measured data, which helps to mitigate this issue.

Table 5. Peak signal-to-noise ratio (PSNR) and structural similarity (SSIM) values averaged across all experiments. Data is aggregated from Tables 1-3. Columns correspond to the averages in each of the three imaging scenarios, as well as the overall average. Rows correspond to the different methods, with the first row being the PSNR value, and the second row the SSIM value. PSNR and SSIM values showing a greater than 2% difference between the proposed method and the next best value are highlighted in bold.

| PSNR (avg) SSIM (avg) | Limited angle (Table 1) | Low-dose (Table 2) | Sparse view (Table 3) | Average |
|---|---|---|---|---|
| Proposed method | **35.7** | 38.3 (-2.0%) | **38.0 (-1.3%)** | **37.3** |
| | **0.978** | **0.977** | **0.982 (-0.2%)** | **0.979** |
| Denoising cycle GAN | 33.9 (-5.0%) | 36.2 (-7.4%) | **38.1 (-1.0%)** | 36.0 (-3.5%) |
| | **0.972 (-0.6%)** | 0.907 (-7.2%) | **0.984** | 0.954 (-2.6%) |
| CIRCLE GAN | 34.2 (-4.2%) | **39.1** | **38.5** | **37.3** |
| | **0.971 (-0.7%)** | **0.977** | **0.982 (-0.2%)** | **0.977 (-0.2%)** |
| SART-TV | 31.6 (-11.5%) | 37.5 (-4.1%) | **38.5** | 35.8 (-4.0%) |
| | **0.966 (-1.2%)** | **0.974 (-0.3%)** | **0.984** | **0.975 (-0.4%)** |
| Pure SART | 30.7 (-14.0%) | 29.1 (-25.6%) | 36.1 (-6.2%) | 32.0 (-14.2%) |
| | 0.932 (-4.7%) | 0.941 (-3.7%) | 0.958 (-2.6%) | 0.944 (-3.6%) |

As shown in Section 3.4, the inclusion of the self-attention block provides some improvement in the quality of the reconstructed images. The concept of self-attention seems promising for CT imaging as the artifacts are often non-local in nature; for example, in low-dose CT, the noise pattern across the image is non-uniform, because it is caused by noisy measurements along lines passing through more attenuating material. Further investigation is required to see if the dependencies modeled by the attention map actually correspond to distinct features in the images.

In the current implementation, the same neural network (the generator of the GAN) is applied between every iteration of SART. The characteristics of the SART image vary considerably as the iteration proceeds, however: in early iterations, the images are blurry as the algorithm is still converging, while in later iterations, the image is sharper, but artifacts or noise may become more pronounced. Thus, it could be advantageous to train the GAN on images produced by intermediate SART iterations, in addition to the final reconstructed images. We may also wish to fine tune other algorithmic parameters such as the number of SART iterations, number of subsets, etc.

In this paper, we have used SART as the iterative reconstruction method to be enhanced. While SART is a well-known reconstruction algorithm for CT, it is a purely algebraic method which does not include any modeling of the noise in the data. It is possible that the use of a statistical method, such as maximum likelihood expectation maximization (MLEM), could therefore provide better results. Additionally, while PSNR and SSIM are standard measures for image quality assessment, they do not necessarily correspond to improved diagnostic potential. A task-based assessment of our method, such as lesion detection, would provide further evidence for the efficacy of the method.

## 5. Conclusion

In this paper, we combine the SART reconstruction algorithm and a Generative Adversarial Network, creating a method to reconstruct CT images from low-dose, sparse view, and limited angle sinogram data. The generator of the GAN is used in between SART iterations as a signal prior, to guide the progress of the reconstruction. In the generator network, we make use of a self-attention block, which helps to recognize long range dependencies in the images. A single GAN is trained on a mixture of training data and applied in a wide range of scenarios, to avoid having to train a network for each specific imaging scenario.

We compared our method with two other neural network based approaches (denoising cycle GAN and CIRCLE GAN), as well as an approach incorporating total variation superiorization (SART-TV). Peak signal to noise ratio (PSNR) and structural similarity index (SSIM) were used to compare the images. The proposed approach provided the best results for limited-angle scenarios, somewhat worse results than cycle GAN and CIRCLE GAN for low-dose imaging, and comparable results for sparse-view imaging. Averaged over all experiments, it provides comparable performance to CIRCLE GAN, and better results than the other methods.

## 6. Acknowledgement

This research was funded by the Graduate Research Award of Computing and Software Systems division, the Startup Fund 74-0525, the Royalty Research Fund 68-2304 of the University of Washington Bothell. We gratefully acknowledge the support of NVIDIA Corporation (Santa Clara, CA, USA) with the donation of the GPU used for this research.